\documentclass{PoS}
\usepackage{bm}
\usepackage{bbm}

\def\gtap{\ \raise.3ex\hbox{$>$\kern-.75em\lower1ex\hbox{$\sim$}}\ }
\def\ltap{\ \raise.3ex\hbox{$<$\kern-.75em\lower1ex\hbox{$\sim$}}\ }
\usepackage{multirow}

\title{Coupled-channel Dalitz plot analysis of $D^+\to K^- \pi^+\pi^+$ decay}

\ShortTitle{Coupled-channel Dalitz plot analysis of $D^+\to K^- \pi^+\pi^+$ decay}

\author{\speaker{Satoshi X. Nakamura}%
\\
Department of Physics,
       Osaka University\\
       E-mail: \email{nakamura@kern.phys.sci.osaka-u.ac.jp}}

\abstract{
We demonstrate that 
partial wave amplitudes extracted from 
$D^+\to K^-\pi^+\pi^+$ Dalitz plot
with a unitary coupled-channel model are significantly different from
those obtained with an isobar model.
The unitary coupled-channel model 
takes account of hadronic rescattering mechanisms involving all three
mesons that have been missed in conventioanl isobar model analyses.
The rescattering mechanisms contribute largely, and can triplicate 
the $D^+\to {K}^-\pi^+\pi^+$ decay width within our analysis.
These findings deliver a warning that analysis results obtained with isobar
models should be looked with a caution.
The determination of the CKM angle $\gamma/\phi_3$ is a highly
relevant problem.
}

\FullConference{VIII International Workshop On Charm Physics\\
                 5-9 September, 2016\\
                 Bologna, Italy}

\begin{document}

\section{Introduction}

\begin{figure}[b]
\begin{center}
 \includegraphics[width=15cm]{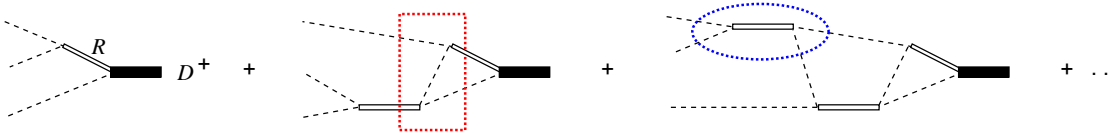}
\end{center}
\caption{\label{fig:d-decay} (Color online)
Diagrammatic representation of $D^+$-decay into three pseudoscalar
mesons in our coupled-channel model.
The dashed lines represent propagations of pseudoscalar mesons.
The part enclosed by 
the red dotted rectangle represents a Z-diagram, while
the blue dotted ellipse cuts out
an off-shell two-pseudoscalar-meson scattering amplitude.
The leftmost diagram is an isobar diagram and the rest are
rescattering diagrams involving the Z-diagrams.
}
\end{figure}

Dalitz plots from heavy meson decays have been conventionally analyzed
with isobar models.
The model assumes, for example, 
a $D$-meson decays into 
an excited state $R$ ($\bar\kappa, \bar K^*$, etc.)
and a pseudoscalar meson, and
the $R$ subsequently decays into a pair of
pseudoscalar mesons;
the third pseudoscalar meson is treated as a spectator 
[Fig.~\ref{fig:d-decay}(left)].
The total decay amplitude is given by a coherent sum of these isobar
amplitudes plus a flat background.
Although the isobar models have been quite successful in getting precise
fits to the data, this does not necessarily mean that 
a (partial wave) decay amplitude extracted with the model is the right
one.
This is because what the data can constrain is only the modulus of the
total decay amplitude.
Therefore, one should use a theoretically sound model for extracting 
amplitudes from the data so that unnecessary model artifact does not
come into play.
An obvious concern about the isobar model is that it misses mechanisms
involving what we call Z-diagrams as shown in
Fig.~\ref{fig:d-decay} (red dotted rectangle).
The Z-diagrams must be considered to take care of full
channel-couplings and the three-body unitarity.

In this contribution, we demonstrate that partial wave amplitudes
extracted from $D^+\to K^-\pi^+\pi^+$ Dalitz plot
with an isobar model are significantly different from those
obtained with a unitary coupled-channel model.
We also show a large contribution from the three-body rescattering
to the Dalitz plot distribution.
Through these investigations, we address
the validity of 
the isobar model in extracting decay amplitudes.
A full account of this work is given in Ref.~\cite{d-decay}.
To the best of our knowledge,
this is the first full Dalitz plot analysis of a $D$-meson decay into three
pseudoscalar mesons with a coupled-channel framework.
Findings here are also relevant to extracting
the CKM angle $\gamma/\phi_3$ from data of
$B^\pm\to \tilde D^0 h^\pm$  ($\tilde D^0=D^0\ {\rm or}\ \bar D^0$; $h=\pi, K$, etc.)
followed by
$\tilde D^0\to$ 3 mesons.
In there, it is crucial to accurately determine the $\tilde D^0$ decay amplitude
and particularly its phase~\cite{gamma}.

The three-body final state interaction in the $D^+\to K^-\pi^+\pi^+$ decay
was also studied and its importance was shown by other groups~\cite{usp,bonn}.
The $K^-\pi^+$ $s$-wave amplitude from an experimental analysis was
studied in Ref.~\cite{usp}.
Meanwhile, the elastic region of the $D^+\to K^-\pi^+\pi^+$ decay Dalitz
plot was analyzed with a dispersion theoretical framework in
Ref.~\cite{bonn}, and decay amplitudes were extracted.

In what follows.
our coupled-channel model to describe
the $D^+\to K^-\pi^+\pi^+$ decay amplitude
is discussed in Sec.~\ref{sec:formalism}.
Then in Sec.~\ref{sec:results},
we present numerical results from our analyses 
of the two-pseudoscalar-meson scattering data, and of
the $D^+\to K^-\pi^+\pi^+$ decay Dalitz plot pseudo-data;
the pseudo-data are generated with the E791 isobar model~\cite{e791} and
thus reasonably realistic.

\section{Formulation}
\label{sec:formalism}

The unitary coupled-channel framework we work with here has been
developed in our previous publications~\cite{3pi-1}.
In this formalism, 
$D^+$-meson decays into a $R h$ channel ($R$: bare resonance), followed by multiple scatterings due
to the hadronic dynamics, leading to the final $K^-\pi^+\pi^+$ state
(Fig.~\ref{fig:d-decay}); both the two-body and three-body unitarity are maintained.
The main driving force for the hadronic rescattering is interactions
between two pseudoscalar mesons.
Therefore, we first construct a two-pseudoscalar-meson ($\pi$, $K$) interaction model that
is subsequently applied to three-pseudoscalar-meson scattering and then the $D$
decay amplitude.
The following presentations are also given in this order.

We describe two-pseudoscalar-meson scatterings with
a unitary coupled-channel model.
For example, we consider $\pi\pi$-$K\bar K$ coupled-channels for a $\pi\pi$
scattering, while $\pi\bar K$-$\eta' \bar K$ coupled-channels for
$I$=1/2 $s$-wave $\pi\bar K$ scattering. 
We model the two-meson interactions with
bare resonance($R$)-excitation mechanisms or contact interactions or both.
These interactions are plugged in Lippmann-Schwinger equation that produces
meson-meson scattering amplitudes.

For a three pseudoscalar-meson scattering, 
let us first assume that the three mesons interact with each other only
through the above-described two-meson interactions.
Because the two-meson interaction is given in a separable
form,
we can cast Faddeev equation into a two-body like
scattering equation (the so-called Alt-Grassberger-Sandhas (AGS) equation~\cite{AGS})
for a ${\cal R}h\to {\cal R}'h'$ scattering.
Here, ${\cal R}$ stands for either $R$ or $r_{ab}$, and $r_{ab}$ is
a spurious ``state'' that is supposed to live within a contact
interaction in a very short time, and decays
into the two pseudoscalar-mesons, $ab$.
The driving force for the scattering is the
Z-diagrams [Fig.~\ref{fig:d-decay} (red dotted rectangle)] 
and the dressed 
${\cal R}$ propagators which include multiple insertions of self energy diagrams. 
In a three-meson system, there may be a room for a 
three-meson-force to play a role.
The hidden local symmetry (HLS) model~\cite{hls} can provide 
vector-meson exchange diagrams that work as a
three-meson-force, and 
we consider them
in our analysis of
the $D^+\to K^-\pi^+\pi^+$ decay to study their relevance.

Finally, 
the decay amplitude for $D^+\to K^-\pi^+\pi^+$
in our coupled-channel model
is diagrammatically represented in Fig.~\ref{fig:d-decay}.
The first term corresponds to the isobar contribution 
while the rests are the contribution from
the hadronic rescattering described by the AGS equation.

\section{Analysis results}
\label{sec:results}

Now we apply the coupled-channel formalism discussed in the previous section to 
analyses of data.
First we determine the two-pseudoscalar-meson
scattering model by analyzing $\pi \bar K$ and $\pi\pi$ scattering data.
Then we analyze 
the $D^+\to K^-\pi^+\pi^+$ decay Dalitz plot.

\subsection{Two-pseudoscalar-meson scattering}
\label{sec:two-meson}

We determine the model parameters
of our $\pi\pi$ and $\pi\bar{K}$ scattering models
by fitting empirical scattering amplitudes for $E\ltap 2$~GeV.
This is the energy region relevant to 
the $D^+\to K^-\pi^+\pi^+$ Dalitz plot analysis.
These two-meson interaction models are a basic ingredient for
describing the three-meson scattering in
the $D^+\to K^-\pi^+\pi^+$ decay.

We analyze the $\pi\bar{K}$ scattering amplitudes from the LASS
experiment~\cite{lass} to determine the model parameters for 
$\{L,I\}$ = \{0,1/2\}, \{0,3/2\}, \{1,1/2\}, \{2,1/2\}
partial waves
($L$: total angular momentum; $I$: total isospin).
For the $\{L,I\}$=\{0,1/2\} wave,
we consider $\pi\bar{K}$-$\eta'\bar{K}$ coupled channels.
For the $\{L,I\}$=\{1,1/2\} and \{2,1/2\} waves,
we consider coupling of $\pi\bar{K}$ and effective inelastic
channels.
Regarding the $\pi\pi$ model, the $\{L,I\}$=\{1,1\},\{0,2\} partial
waves are needed for our coupled-channel analysis of the $D^+\to K^-\pi^+\pi^+$ decay.
We consider $\pi\pi$-$K\bar{K}$ coupled channels for 
for $\{L,I\}$=\{1,1\} and 
the elastic $\pi\pi$ channel for $\{L,I\}$=\{0,2\},
and fit the CERN-Munich data~\cite{hyams}.

\begin{figure}[t]
\includegraphics[width=0.33\textwidth]{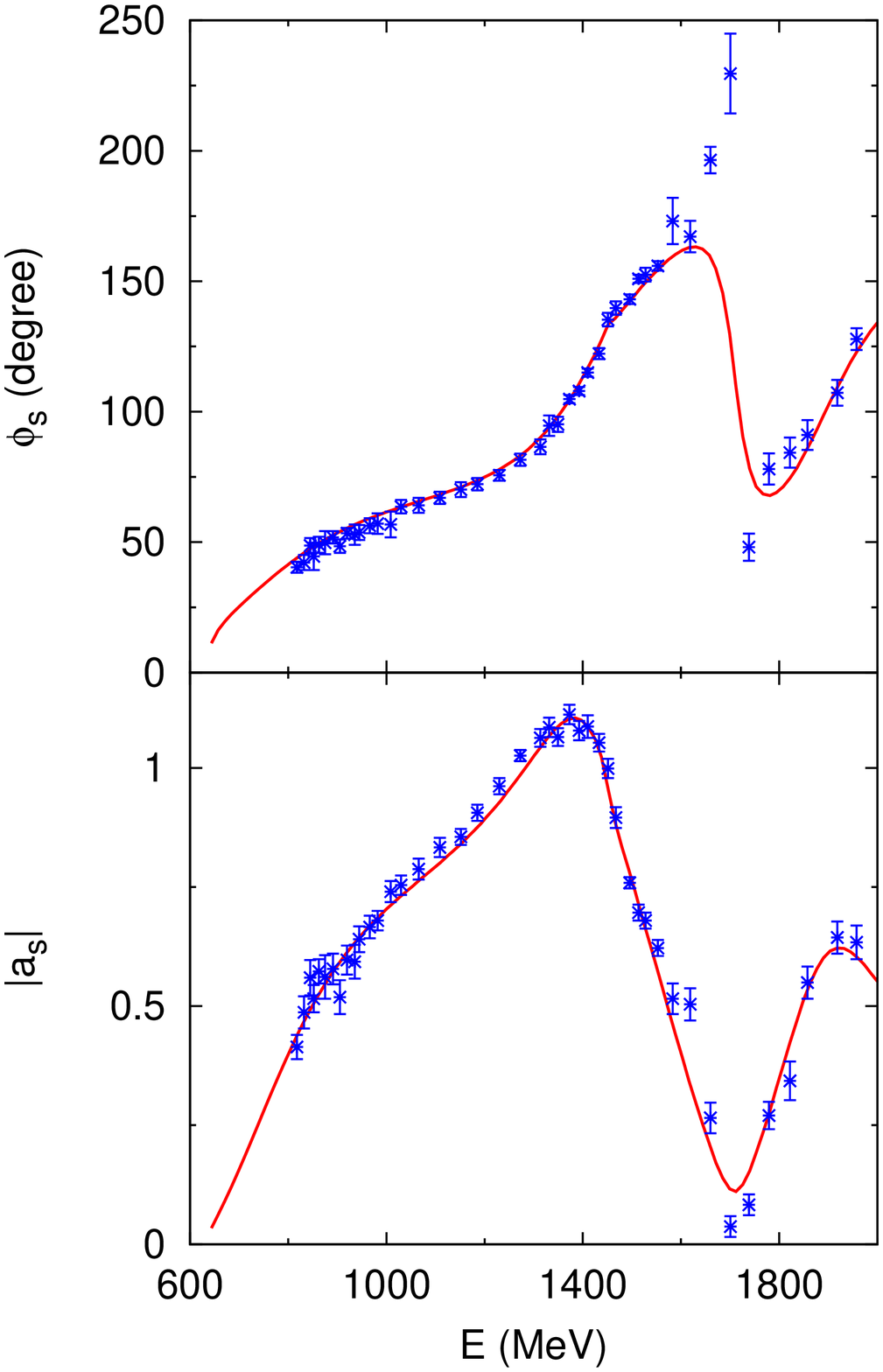}
\hspace{-2mm}
\includegraphics[width=0.33\textwidth]{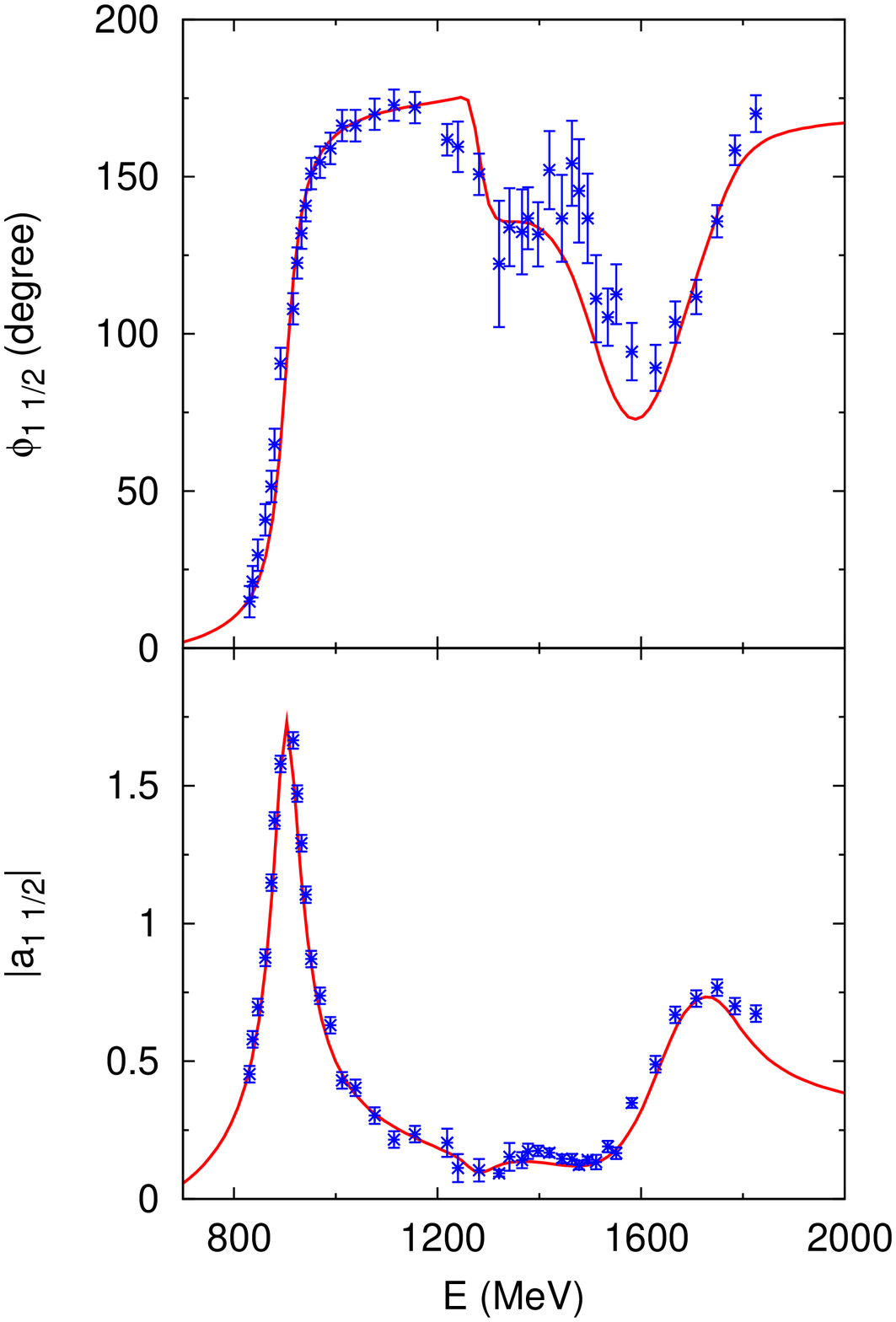}
\hspace{-2mm}
\includegraphics[width=0.33\textwidth]{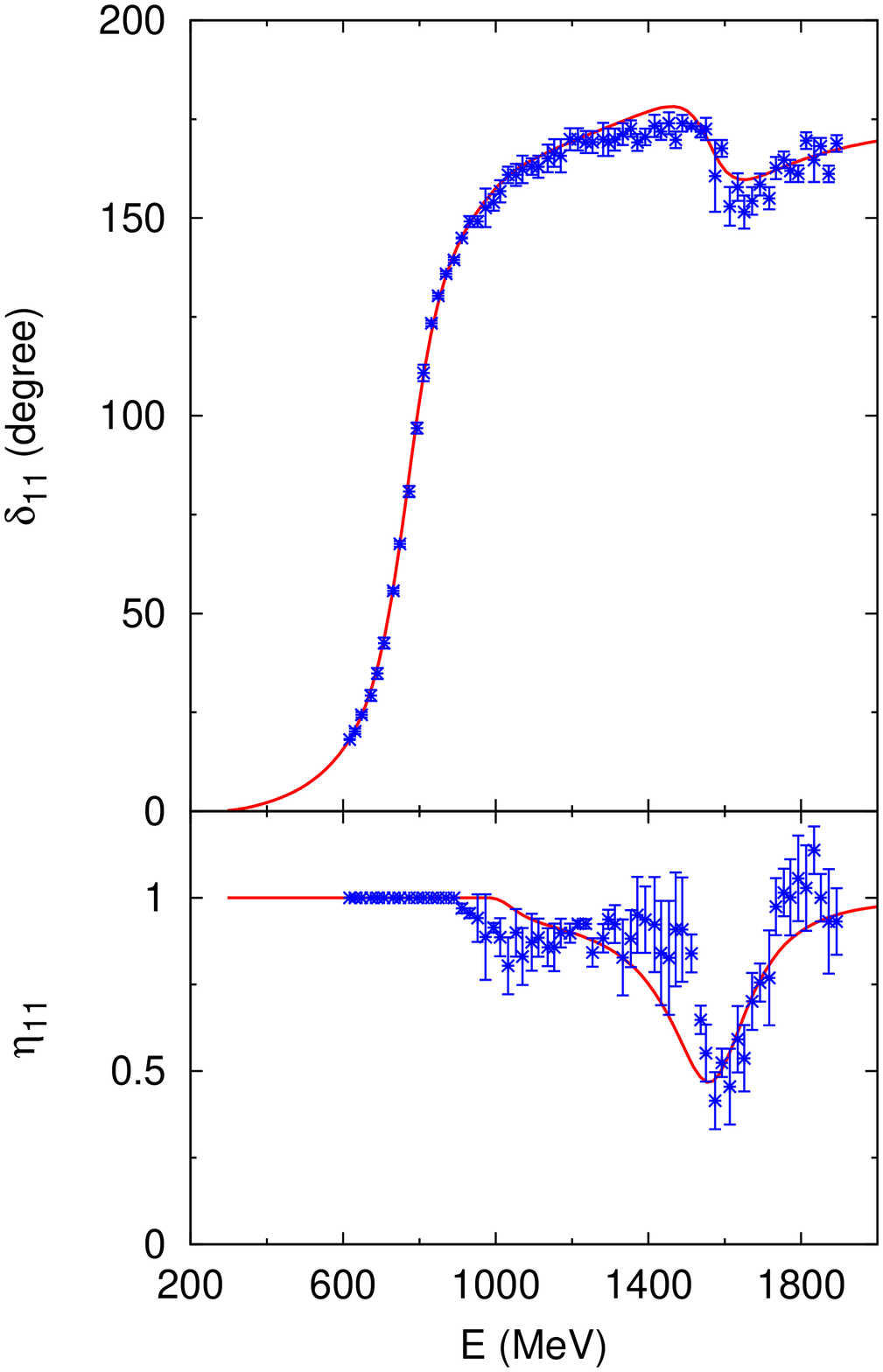}
\caption{\label{fig:pik} 
(Color online) 
Phase (upper) and modulus (lower) of 
the $\pi\bar{K}$ scatterings amplitudes:
(Left) $L$=0 for $\pi^+K^-$ ; (Center) $\{L,I\}$=\{1,1/2\}; 
Data are from Ref.~\cite{lass}.
(Right) Phase-shifts and inelasticities for $\{L,I\}$=\{1,1\} $\pi\pi$ scattering.
Data are from Ref.~\cite{hyams}.
Figures taken from Ref.~\cite{d-decay}. Copyright (2016) APS.
}
\end{figure}
We present the quality of the fits to the empirical
amplitudes for the $\pi^+K^-$ $L$=0 and 
$\{L,I\}$=\{1,1/2\} partial waves in Fig.~\ref{fig:pik} (left, center)
where phases 
(upper panels) and modulus (lower panels) of the amplitudes are shown.
For the $\pi\pi$ scattering, we present
phase shifts and inelasticities for
the $\{L,I\}$=\{1,1\} wave in
Fig.~\ref{fig:pik} (right).
As seen in the figure,
the $\pi\pi$ and $\pi\bar K$ amplitudes our model generates are
reasonable, and should be good enough
for a coupled-channel analysis of the $D^+ \to K^- \pi^+ \pi^+$.

\subsection{Analysis of $D^+\to K^-\pi^+\pi^+$ Dalitz plot}
\label{sec:D-decay}

Now we perform a partial wave analysis of 
the $D^+\to K^-\pi^+\pi^+$ Dalitz plot pseudo-data 
using our coupled-channel model.
We explain setups of our models used in the analysis.
In our coupled-channel framework,
$D^+$-meson decays into ${\cal R}h$ channels, and then hadronic
rescatterings follows before a final $K^-\pi^+\pi^+$ state appears (Fig.~\ref{fig:d-decay}).
We consider the following 11 ${\cal R}h$ coupled channels in our full calculation:
\begin{eqnarray}
\{{\cal R}h\} = \left\{
R^{01}_1\pi,
R^{01}_2\pi,
r^{01}_{\pi\bar K}\pi,
R^{11}_1\pi,
R^{11}_2\pi,
R^{11}_3\pi,
R^{21}_1\pi,
R^{12}_1\bar{K},
R^{12}_2\bar{K},
r^{03}_{\pi\bar K}\pi,
r^{04}_{\pi\pi}\bar{K}
\right\} \ ,
\label{eq:cc}
\end{eqnarray}
where 
$R^{L,2I}_i$ stands for $i$-th bare $R$ state with the spin $L$ and the
 isospin $I$; when $I$ is an integer (half-integer), it is understood
 that this $R$ state has the strangeness $S$=0 ($S$=$-1$) in this report.
Thus, $R^{01}_i$, $R^{11}_i$, $R^{21}_i$, $R^{12}_i$ are seeds of
$\bar K^*_0$, $\bar K^*$, $\bar K^*_2$, $\rho$ resonances, respectively.
In our model, these resonances are included as poles in the unitary
scattering amplitudes. 
The $\pi\pi$ $\{L,I\}$=\{1,1\} partial wave associated with the
$R^{12}_i\bar{K}$ channel has not been considered in the previous
isobar analyses of $D^+\to K^-\pi^+\pi^+$.
This channel can contribute 
only through channel-couplings induced by the Z-diagrams,
and therefore it does not show up in isobar models.
We do not include a flat interfering background amplitude.
We fit Dalitz plot pseudo-data for $D^+\to K^-\pi^+\pi^+$ by 
adjusting parameters associated with 
the $D^+\to {\cal R}h$ vertices (couplings, phases).

In our analysis of $D^+\to K^-\pi^+\pi^+$ Dalitz plot pseudo-data,
three models are used.
The first one is the ``Full model'' that contains all the dynamical contents
described above.
The second one is the ``Z model'' for which the rescattering mechanism is
solely due to multiple iteration of the two-pseudoscalar-meson interactions 
in the form of the Z diagrams and ${\cal R}$ propagators; no three-meson-force.
The third model is the ``Isobar model'' that does not explicitly include
the Z-diagrams.
Finally we remark that the two-pseudoscalar-meson interactions
will not be adjusted to fit the $D^+$-decay pseudo-data.
This is in contrast with most of the previous analyses where some of
Breit-Wigner parameters associated with $R$-propagations
were also adjusted along with $D^+\to Rh$ vertices.
After the fits, we obtained the Full, Z, and Isobar models
for which $\chi^2/{\rm d.o.f.}$ = 0.22, 0.17, and 0.42, respectively.

\begin{figure}[t]
\includegraphics[width=0.33\textwidth]{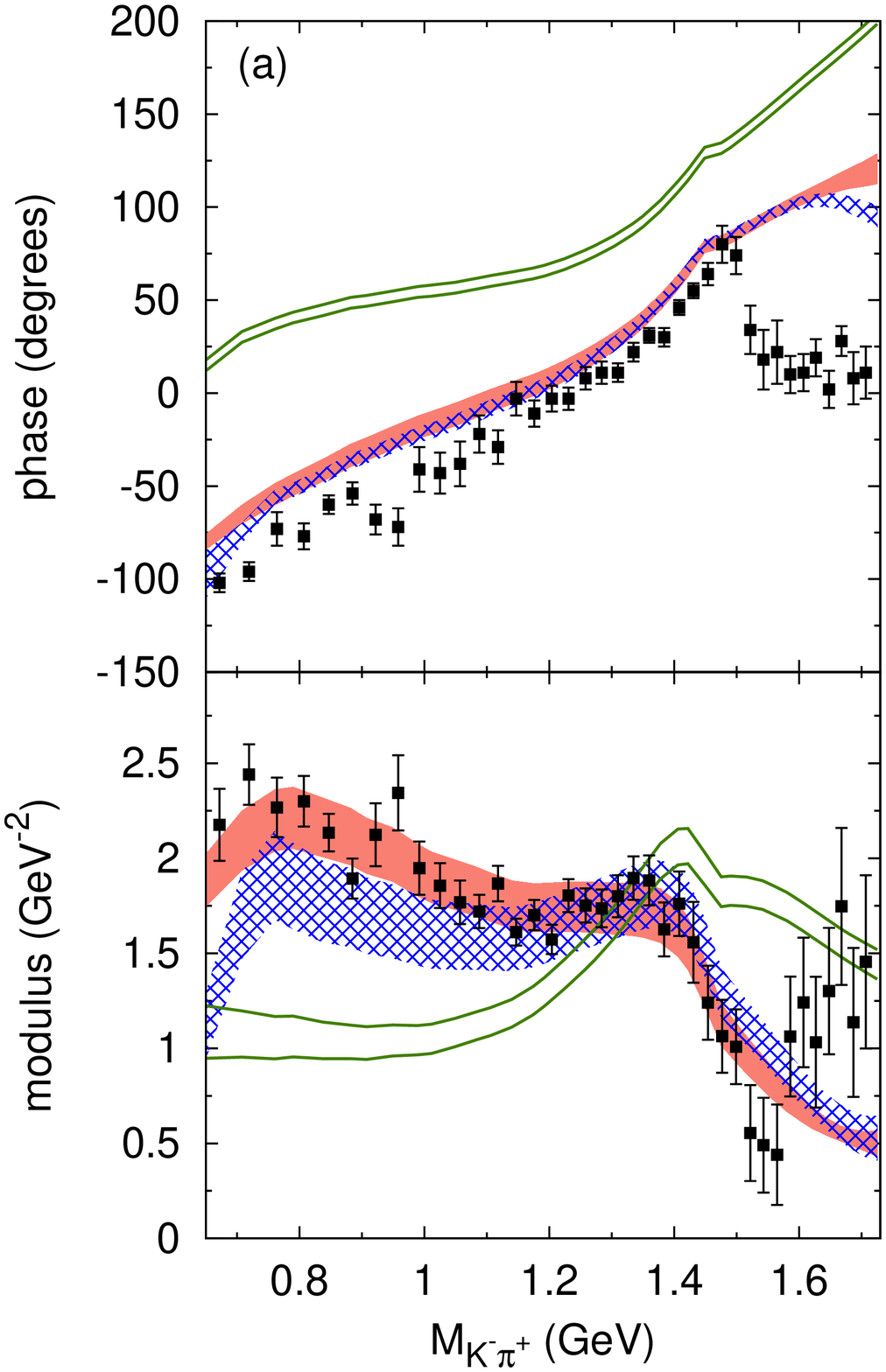}
\hspace{-3mm}
\includegraphics[width=0.33\textwidth]{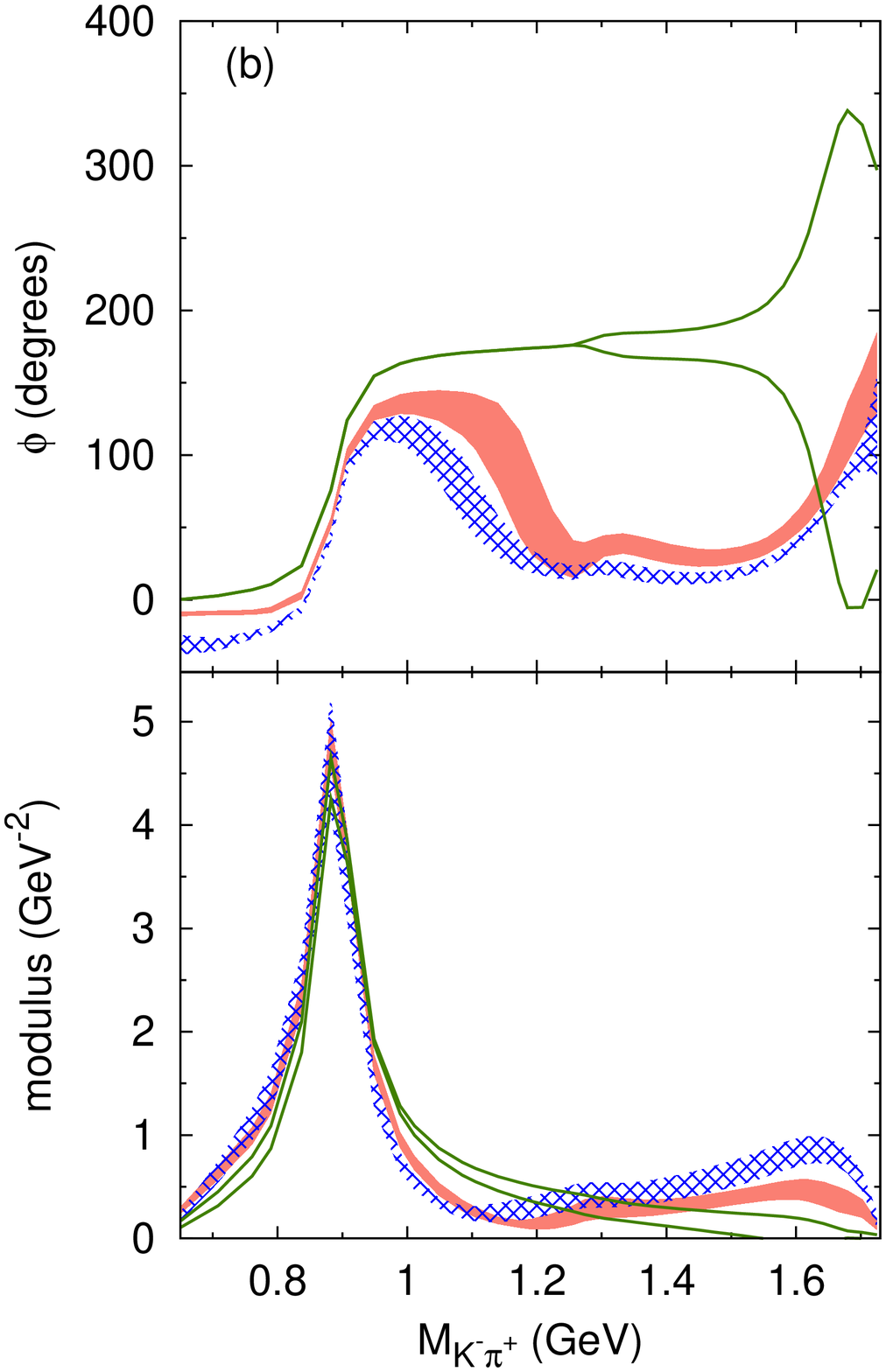}
\hspace{-1mm}
\includegraphics[width=0.33\textwidth]{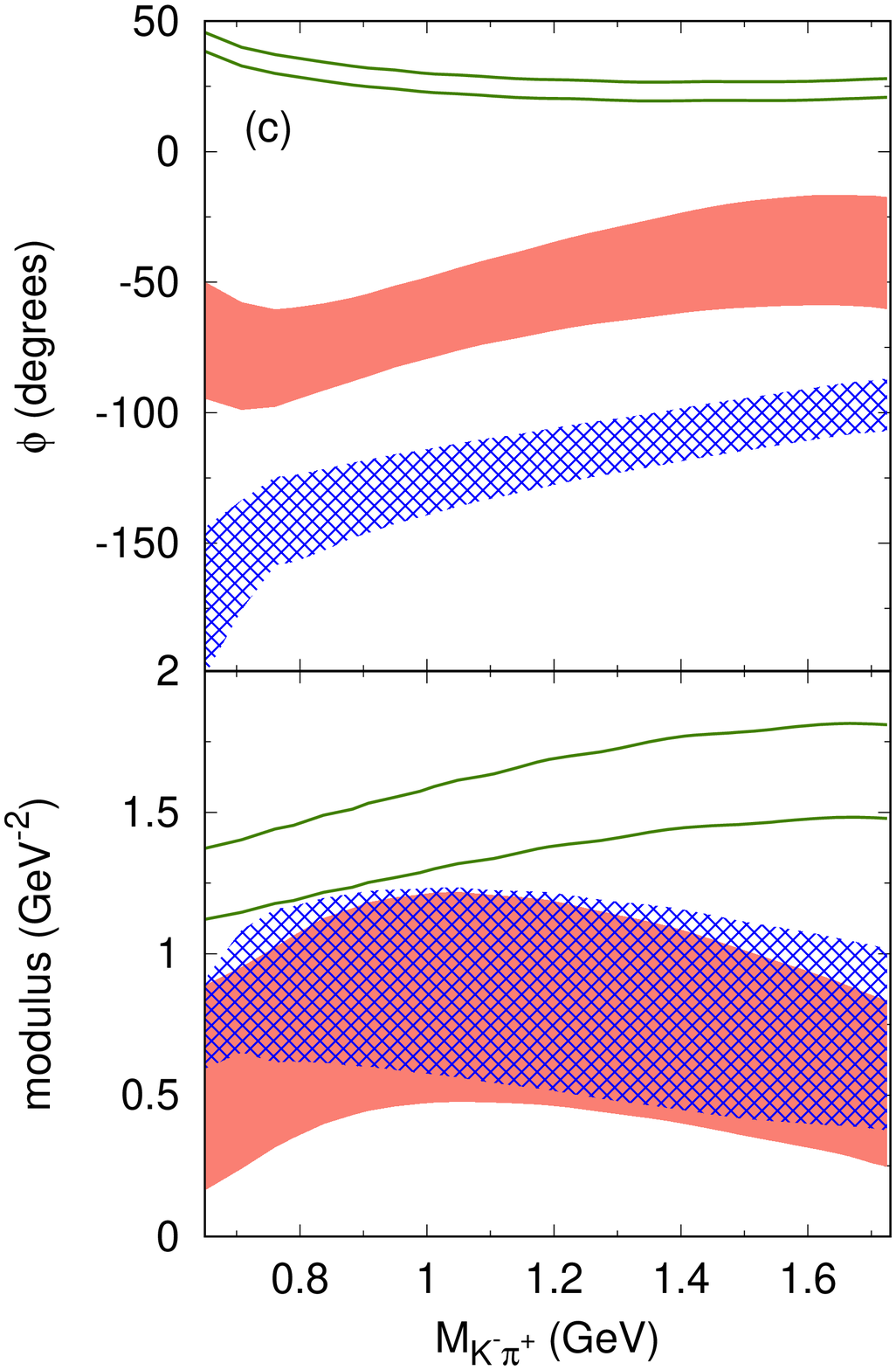}
\caption{\label{fig:amp-err} 
(Color online) 
Partial wave amplitudes extracted from the $D^+\to K^-\pi^+\pi^+$ Dalitz plot.
Phase (upper) and modulus (lower) are shown for
(a) $(K^-\pi^+)_S^{I=1/2}\pi^+$; 
(b) $(K^-\pi^+)_P^{I=1/2}\pi^+$;
(c) $(K^-\pi^+)_S^{I=3/2}\pi^+$.
The red solid, blue cross-hatched, and green bordered
bands are for the Full, Z, and Isobar models, respectively;
the band widths represent the statistical errors. 
The $(K^-\pi^+)_S\pi^+$ amplitude from
the MIPWA of the E791 Collaboration~\cite{e791} is shown by
the black squares.
Figures taken from Ref.~\cite{d-decay}. Copyright (2016) APS.
}
\end{figure}
We present in Fig.~\ref{fig:amp-err}
partial wave decay amplitudes extracted in the analysis;
the horizontal axis is the invariant mass of $K^-\pi^+$, $M_{K^-\pi^+}$.
We denote a partial wave by ``$(ab)^I_{L}c$'' in which
a two-pseudoscalar-meson pair $ab$ has the total angular momentum $L$ 
and the total isospin $I$.
The partial wave amplitudes for 
$(K^-\pi^+)_S^{I=1/2}\pi^+$,
$(K^-\pi^+)_P^{I=1/2}\pi^+$, and 
$(K^-\pi^+)_S^{I=3/2}\pi^+$
are presented in Figs.~\ref{fig:amp-err}(a)-(c), respectively.
As a whole, the Full and Z models are similar in the amplitudes
while the Isobar model is rather different, particularly in 
$(K^-\pi^+)_S^{I=1/2}\pi^+$ and
$(K^-\pi^+)_S^{I=3/2}\pi^+$.
Because the models maintain the Watson theorem
when the Z-diagrams are absent,
the phases (the upper panels of Fig.~\ref{fig:amp-err})
from the Isobar model in the elastic region are essentially the
same as those of the LASS $\pi\bar K$ amplitude
up to overall constant shifts.
The difference in the $M_{K^-\pi^+}$-dependence of 
the phases between the Isobar
model and the Full (or Z) model is solely due to the effect of the hadronic
rescattering involving the Z-diagrams. 

In Fig.~\ref{fig:amp-err}(a), we also show the 
$(K^-\pi^+)_S\pi^+$ amplitude from the E791 MIPWA
(model independent partial wave analysis)~\cite{e791}
denoted by 
$[(K^-\pi^+)_S\pi^+]^{\rm\scriptscriptstyle E791}_{\rm\scriptscriptstyle MIPWA}$. 
Interestingly, the $M_{K^-\pi^+}$-dependence of 
the phases from the Full and Z models are in a very good
agreement with those of the MIPWA for $M_{K^-\pi^+}\ltap 1.5$~GeV.
Thus, our coupled-channel models explain
the difference between the phase of 
$[(K^-\pi^+)_S\pi^+]^{\rm\scriptscriptstyle E791}_{\rm\scriptscriptstyle MIPWA}$
and that of
the LASS $(K^-\pi^+)_S^{I=1/2}\pi^+$ amplitude
in a way qualitatively different from the previous explanation.
Edera et al.~\cite{pik_I=3/2_model} and the FOCUS $K$-matrix model
analysis~\cite{focus} explained the difference with a rather large (more than 100\%)
destructive interference between 
the $(K^-\pi^+)_S^{I=1/2}\pi^+$ and $(K^-\pi^+)_S^{I=3/2}\pi^+$ 
amplitudes, without an explicit consideration of the three-body hadronic rescattering. 
Our coupled-channel models, on the other hand, explained the difference with the hadronic
rescattering involving the Z-diagrams, and have a moderately 
destructive interference between 
the $(K^-\pi^+)_S^{I=1/2}\pi^+$ and $(K^-\pi^+)_S^{I=3/2}\pi^+$
amplitudes.

We notice in Fig.~\ref{fig:amp-err}(c) that
the $(K^-\pi^+)_S^{I=3/2}\pi^+$
partial wave amplitude has relatively large errors.
Because we analyzed the data of the single charge state,
it may be difficult to separate the different isospin states 
with a good precision. 
This situation would be improved by analyzing data of different charge
states, i.e., $D^+\to K^-\pi^+\pi^+$ and $D^+\to K_S^0\pi^+\pi^0$,
in a combined manner.

\begin{figure}[t]
\includegraphics[width=0.49\textwidth]{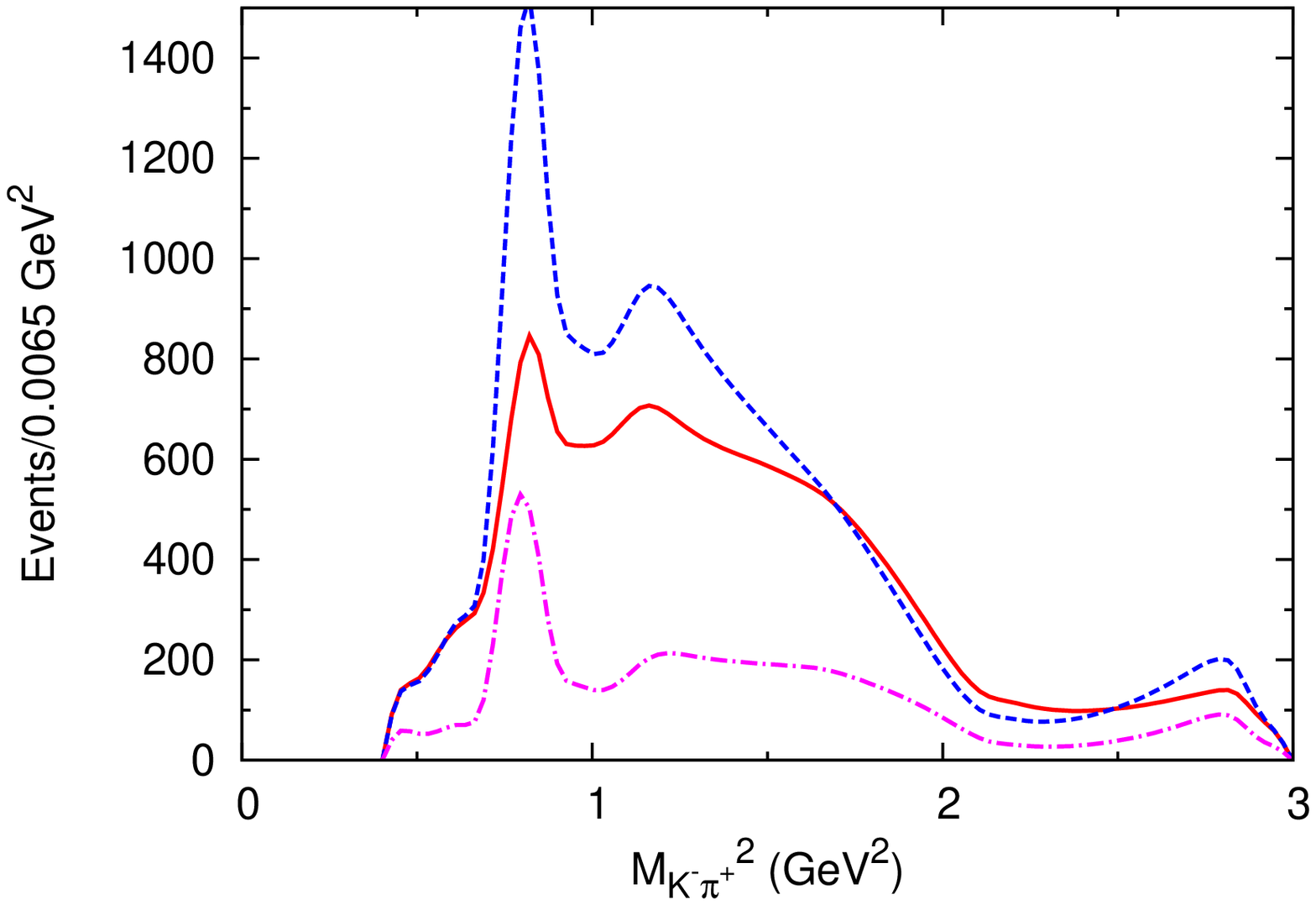}
\hspace{5mm}
\includegraphics[width=0.49\textwidth]{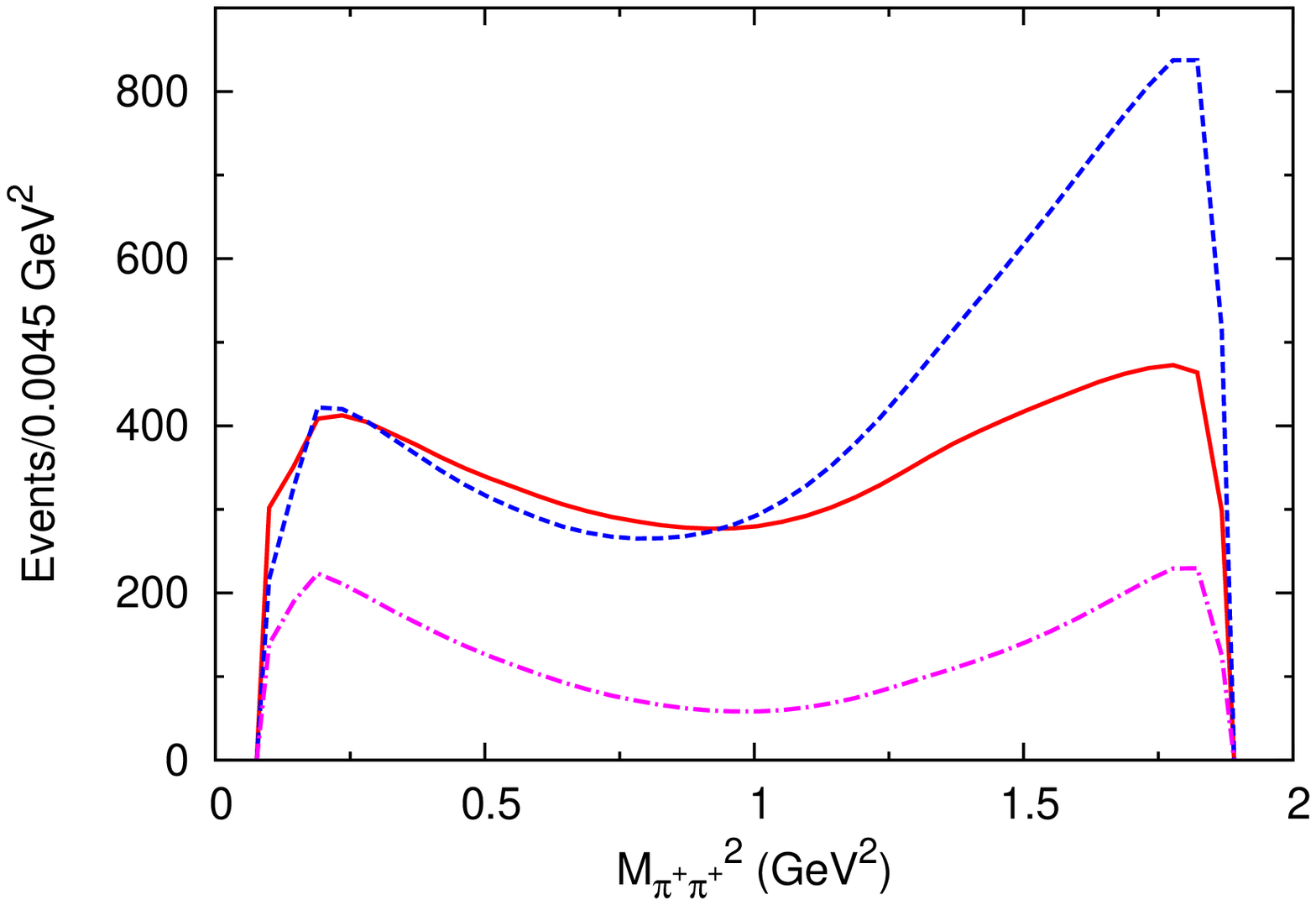}
\caption{\label{fig:3mf}
(Color online) 
The $K^-\pi^+$(left) and $\pi^+\pi^+$(right) squared invariant mass spectrum
for $D^+\to K^-\pi^+\pi^+$.
The red solid curves are from the Full model.
The blue dashed curves are also from the Full model but
the three-meson-force is turned off.
The magenta dot-dashed curves are from the Full model with
all the rescattering involving the Z-diagrams turned off.
Figures taken from Ref.~\cite{d-decay}. Copyright (2016) APS.
}
\end{figure}
Now let us study how much the three-meson-force contributes
to the $D^+\to K^-\pi^+\pi^+$ decay.
In Fig.~\ref{fig:3mf}, we compare 
the $K^-\pi^+$ ($\pi^+\pi^+$) squared invariant mass spectrum of the Full model
with those from the same model but the three-meson-force being turned
off. 
The three-meson-force suppresses
the decay width by $\sim$22\%, and 
change the spectrum shape significantly, as seen in Fig.~\ref{fig:3mf}.
In the same figure, we also show the spectrum from the Full model with
all the Z-diagrams
(diagrams except for the first term in Fig.~\ref{fig:d-decay})
being turned off.
Effects of the Z-induced rescattering mechanisms are quite large;
the decay width gets almost triplicated by the rescattering effect.
In particular, we found that 
the hadronic rescattering through partial wave
$(\pi^+\pi^0)_P^{I=1}\bar{K}^0$, which includes $\rho(770)$, gives a
major contribution.

\section{Summary and future perspective}
\label{sec:summary}

The analysis conducted here clearly showed the importance of the
hadronic rescattering involving the Z-diagrams that have been missed in
the conventional isobar model analyses. 
Partial wave amplitudes
extracted from Dalitz data with an isobar model should be looked with a
caution.

For a further improved analysis, 
a combined analysis of related processes would greatly help.
For example, by simultaneously analyzing
the $D^+\to K^-\pi^+\pi^+$ and $D^+\to K_S^0\pi^0\pi^+$ decays,
the isospin separation of the amplitude would be better done.
Also, we expect the role of $\rho(770)$ in the decays to be extracted
reliably, because the BESIII analysis~\cite{bes3} found
its dominant contribution to $D^+\to K_S^0\pi^0\pi^+$.
We stress that combined analyses are very common for the
baryon spectroscopy~\cite{knls13}, and the coupled-channel
unitarity is a guiding principle there. 
In the CHARM 2016 workshop, we heard three talks on $D^+\to K^-K^+K^+$
Dalitz plot analysis~\cite{alberto,pat}.
The experimental analysis found three comparable solutions for the
$K^-K^+$ $s$-wave amplitude~\cite{alberto}.
A combined analysis of $D^+\to K^-K^+K^+$ and $K^+\pi^-\pi^+$
could discriminate one from the other solutions.

\begin{acknowledgments}
\vspace{-2mm}
My participation to CHARM 2016 was supported by JSPS KAKENHI Grant Number JP25105010.
\end{acknowledgments}


\vspace{-2mm}
\begin{thebibliography}{99}

\bibitem{d-decay}
S.X. Nakamura, {\it Phys. Rev. D} {\bf 93} (2016) 014005. 

\bibitem{gamma}
A. Poluektov et al. (Belle Collaboration),
{\it Phys. Rev. D} {\bf 81} (2010) 112002. 

\bibitem{usp}
P.C. Magalh\~aes et al.,
{\it Phys. Rev. D} {\bf 84} (2011) 094001;\\
P.C. Magalh\~aes and M.R. Robilotta, 
{\it Phys. Rev. D} {\bf 92} (2015) 094005. 

\bibitem{bonn}
F. Niecknig and B. Kubis, {\it JHEP} {\bf 1510} (2015) 142.

\bibitem{e791}
E.M. Aitala et al. (E791 Collaboration),
{\it Phys. Rev. D} {\bf 73}  (2006) 032004. 

\bibitem{3pi-1}
H. Kamano, S.X. Nakamura, T.-S.H. Lee, and T. Sato,
{\it Phys. Rev. D} {\bf 84} (2011) 114019;\\
S.X. Nakamura, H. Kamano, T.-S.H. Lee, and T. Sato,
{\it Phys. Rev. D} {\bf 86} (2012) 114012.

\bibitem{AGS}
E.O. Alt, P. Grassberger, and W. Sandhas, {\it Nucl. Phys. B} {\bf 2} (1967) 167.

\bibitem{hls}
M. Bando, T. Kugo, and K. Yamawaki,
{\it Phys. Rept.} {\bf 164} (1988) 217.

\bibitem{lass}
D. Aston et al., {\it Nucl. Phys. B} {\bf 296} (1988) 493;\\
P. Estabrooks et al., {\it Nucl. Phys. B} {\bf 133} (1978) 490.

\bibitem{hyams}
B. Hyams et al., {\it Nucl. Phys. B} {\bf 64} (1973) 134.

\bibitem{pik_I=3/2_model}
L. Edera and M.R. Pennington,
{\it Phys. Lett. B} {\bf 623} (2005) 55.

\bibitem{focus}
J.M. Link et al. (FOCUS Collaboration),
{\it Phys. Lett. B} {\bf 653} (2007) 1.

\bibitem{bes3}
M. Ablikim et al. (BESIII Collaboration),
{\it Phys. Rev. D} {\bf 89} (2014) 052001.

\bibitem{knls13}
H. Kamano, S.X. Nakamura, T.-S.H. Lee, and T. Sato,
{\it Phys. Rev. C} {\bf 88} (2013) 035209.


\bibitem{alberto}
A.C. dos Reis, {\it this proceedings};\\
T. Evans, {\it this proceedings}.\\

\bibitem{pat}
P.C. Magalh\~aes, {\it this proceedings};\\
R.T. Aoude, P.C. Magalh\~aes, A.C. dos Reis, and M.R. Robilotta, arXiv:1604.02904.

\end{thebibliography}
\end{document}